# Reorganization energy from charge transport measurements in a monolithically-integrated molecular device

Leandro Merces,[1] Graziâni Candiotto,[2,3] Letícia M. M. Ferro,[1,4] Anerise de Barros,[4] Carlos V. S. Batista,[1,5] Ali Nawaz,[6] Antonio Riul Jr,[7] Rodrigo B. Capaz,[1,2] and Carlos C. Bof Bufon[1,4,5],*

[1]Brazilian Nanotechnology National Laboratory (LNNano), Brazilian Center for Research in Energy and Materials (CNPEM), 13083-100 Campinas SP, SP, Brazil.
[2]Instituto de Física, Universidade Federal do Rio de Janeiro (UFRJ), 21941-972 Rio de Janeiro RJ, Brazil.
[3]Instituto de Química, Universidade Federal do Rio de Janeiro (UFRJ), 21941-909 Rio de Janeiro RJ, Brazil.
[4]Institute of Chemistry (IQ), University of Campinas (UNICAMP), 13083-970 Campinas SP, Brazil.
[5]Postgraduate Program in Materials Science and Technology (POSMAT), São Paulo State University (UNESP), 17033-360 Bauru SP, Brazil.
[6]Bruno Kessler Foundation (FBK), 38123 Trento, Italy.
[7]Department of Applied Physics, "Gleb Wataghin" Institute of Physics (IFGW), University of Campinas (UNICAMP), 13083-859 Campinas SP, Brazil.
*corresponding author: carlos.bufon@Lnnano.cnpem.br

**Abstract:** Intermolecular charge transfer reactions are key processes in physical chemistry. The electron-transfer rates depend on a few system's parameters, such as temperature, electromagnetic field, distance between adsorbates and, especially, the molecular reorganization energy. This microscopic greatness is the energetic cost to rearrange each single-molecule and its surrounding environment when a charge is transferred. Reorganization energies are measured by electrochemistry and spectroscopy techniques as well as at the single-molecule limit using atomic force microscopy approaches, but not from temperature-dependent charge transport measurements nor in a monolithically-integrated molecular device. Nowadays self-rolling nanomembrane (rNM) devices, with strain-engineered mechanical properties, on-a-chip monolithic integration, and operable in distinct environments, overcome those challenges. Here, we investigate the charge transfer reactions occurring within a *ca.* 6 nm thick copper-phthalocyanine (CuPc) film employed as electrode-spacer in a monolithically integrated nanocapacitor. Employing the rNM technology allows us to measure the molecules' charge-transport dependence on temperature for different electric fields. Thereby, the CuPc reorganization energy is determined as $(930 \pm 40)$ meV, whereas density functional theory (DFT) calculations support our findings with the atomistic picture of the CuPc charge transfer reaction. Our approach presents a consistent route towards electron transfer reaction characterization using current-voltage spectroscopy and provides insight into the role of the molecular reorganization energy when it comes to electrochemical nanodevices.

# Introduction

The charge transport mechanism in organic materials is often a complex combination of inter- and intramolecular electron transfer processes.[1–4] Charge transfer rates for intramolecular transport are expected to be much higher than those provided by intermolecular conduction, as shown in recent experimental[5–7] and theoretical[8] studies. Accordingly, it is a consensus in the literature that intramolecular charge transport must be optimized in targeting high-efficiency organic semiconductor applications, *e.g.*, solar cells,[9,10] light-emitting diodes,[11,12] transistors,[13–15] and spin valves.[16] However, it is extremely hard to assess experimentally the microscopic parameters responsible for intermolecular charge transfer, as both relaxation and decoherence play fundamental roles in quantum mechanical tunneling when varying the transport distance ($d$) and the reservoir temperature ($T$).[17–19]

As intermolecular redox reactions critically affect electrochemistry at the nanoscale, a possible route for obtaining the charge-transfer related microscopic parameters is through well-controlled electronic transport measurements in molecular devices. Down at the molecular level, charge transport mechanisms can broadly be divided into charge transfer kinetics *via* decoherence (hopping) and regimes dominated by coherent tunneling.[20] The former is well described by the Marcus formalism,[21] whereas the latter derives from the Landauer theory.[22] The conceptual difference between them is that Marcus theory describes thermally-activated transport over an energy barrier, whereas the Landauer formalism essentially describes $T$-independent tunneling across the heterostructure.[23,24] In practice, the Marcus $T$-dependent processes differ from the Landauer mechanisms in allowing full charge-carrier relaxation combined with nuclear reorganization within the molecules and their surrounding units. As we shall see, this defines the reorganization energy $\lambda$, a pivotal parameter in the charge transfer activation energy determination.

The importance of $\lambda$ transcends physical chemistry since electron transfer and energy transduction are entangled in nature. Not by chance, $\lambda$ is defined as the energetic cost to rearrange each single-molecule and its surrounding environment when a redox reaction occurs – *i.e.*, an electron is transferred.[25] As a microscopic physical greatness of molecular systems, $\lambda$ draws the picture of solid-state electronic transport systems, organic and inorganic chemical reactions, photosynthetic redox processes, electron transfer in proteins and other biological media, among others.[26] Consolidate approaches to measure $\lambda$ at different physical scales can be found in the literature. Macroscopically, spectroscopy techniques as well as electrochemistry experiments can be used.[27–31] At the single-molecule limit, atomic force microscopy (AFM) with single-charge sensitivity was employed recently to determine $\lambda$ from molecules measured on insulating substrates.[32] However, the bridge that connects these two physical-chemical cosmos, *viz.* (spectro)electrochemistry and molecular electronics, and provides a self-consistent approach to measure $\lambda$ using a molecular electrochemical device, is not built so far.

Here, we discuss how the Marcus theory can be applied to obtain the molecular reorganization energy from charge transport measurements in a monolithically-integrated nano-sized capacitor (*nCap*). Our device consists of a metal/molecular-ensemble stack vertically connected to a metal self-rolled top-electrode. Such approach avoids electrical short-circuits usually caused by top-electrode evaporation, whereas the mechanically soft top-contact preserves the molecules' integrity and function. Additionally, monolithic integration spares the use of either liquid metal electrodes or scanning probe tips, allowing reliable fabrication of more-likely real-life molecular devices that can be operated in different electromagnetic fields, temperatures, pressures, among other outer conditions.[10,17] In the literature, one may find exciting examples of using the nanomembrane self-rolling (rNM) technology to fabricate monolithically-integrated devices featuring organic/inorganic hybrid functions.[3,13,33–36]

As a validating target for our experimental methodology, we employed ultrathin (< 10 nm) copper-phyhalocyanine (CuPc) films that have been considered as prototypical materials to advance nanotechnology toward new electrochemistry applications.[3,13,33,37,38] The *nCap* component's topography and morphology are evaluated by atomic force microscopy (AFM) and scanning electron microscopy (SEM), whereas the CuPc film surface potential analyzed by Kelvin probe force microscopy (KPFM). Raman spectroscopy characterization is performed to attest of the CuPc integrity as a *nCap* counterpart. The *T*-dependent electrical properties are evaluated by current-voltage (*I-V*) characterization, and the reorganization energy is obtained from charge transport measurements pioneering the use of a monolithically integrated nanodevice. To corroborate our experimental findings, *ab initio* calculations nicely confirm the experimental predictions by providing insightful diagrams for the reorganization energy as a function of the CuPc film's static and optical dielectric constants. The robustness of our methodology paves the way towards electron transfer reaction characterization using simple electrical measurements and provides insight into the role of the molecular reorganization energy when it comes to monolithically integrated electrochemical nanodevices.

## Methods

**Nanodevice Fabrication.** The *nCap*s are fabricated on SiO$_2$-coated (2 μm thick), 9 mm × 9 mm Si (100) substrates. The device manufacturing relies on standard microfabrication procedures, *viz.* optical lithography, etching processes, and thin-film deposition. All metallic layers are deposited using electron beam evaporation at a high vacuum (*ca.* 10$^{-6}$ Torr). The organic material is deposited using a resistive thermal evaporator (Leybold Univex 250 system) at a high vacuum (*ca.* 10$^{-6}$ Torr) while the film thickness (*t*) is monitored by quartz crystal microbalances positioned inside the deposition chambers. During each deposition step, the substrate is kept at room temperature. The *nCaps* fabrication followed the particular steps reported in our previous works.[37,39] The self-rolling process is carried out in a 0.25% (v/v) H$_2$O$_2$ aqueous solution, which selectively removes the GeO$_x$ sacrificial layer and leads to the rNM release from the substrate.[34] The as-fabricated devices were stored in a high vacuum (*ca.* 10$^{-5}$ Torr) for at least 24 h in order to remove any water residue incorporated within the device structure during the fabrication process.

**Electrical Characterization.** The I-V measurements are realized using a Keithley 4200 SCS (Semiconductor Parameter Analyzer). I-V characterization as a function of T is done in a Lakeshore EMPX-HF cryogenic probe station. Electrical characterization is performed for 6 as-fabricated *nCaps*, whereas the error bars arise from their statistical deviation.

**Structure Characterization.** The topography and morphology of the heterostructures are evaluated using a Nanosurf NaioAFM and a FEI Inspect F50, respectively. The KPFM measurements are performed using a Park NX10, operating in tapping mode with a Si cantilever (75 kHz resonance) and a coated tip (Pt/Ir, 25 nm radius). The Raman spectra are acquired in a Horiba XploRA PLUS confocal Raman microscope, with circularly-polarized 638 nm laser, a 100× objective, and a 25% filter. The SEM images are colored using Paint.NET free image editing software. The AFM and KPFM images are analyzed using Gwyddion data analysis software. The graphs are plotted using KaleidaGraph software, whereas no baseline/spike correction is applied in the Raman spectra.

***Ab Initio* Calculation.** Calculations were performed using GAUSSIAN03 code.[40] We used the hybrid functional Becke three-parameter Lee-Yang-Parr (B3LYP)[41,42] along with a split-valence triple-zeta polarized Gaussian type orbital basis, 6-311G (d, p).[43] To probe the effects of the applied voltage in the reorganization energies, calculations were performed with and without an external electric field of 1.0 MV/cm, applied along the plane of the molecule, in consistency with the molecular orientation with respect to the substrate (Figure 1a). The dielectric environment was introduced in the calculations through the polarized continuum model (PCM).

**Results and Discussion**

CuPc thin films are recognized in the literature for their exceptional thermal and chemical stability.[38] One may found that the CuPc growth on crystalline substrates occurs in two crystallographic phases, namely, α and β.[44] The metastable α-CuPc grows if the substrate is kept at room temperature during film deposition. The thermodynamically stable β-CuPc is formed by growing the film on a heated substrate or by post-growth annealing.[44] The CuPc molecules in α-phase are arranged in a slipped-stack pattern,[45] while in β-phase the molecules should pack into a herringbone structure.[8] The primary difference between the two crystallographic phases is the tilt angle of the molecular plane with respect to the substrate.[46] In our case, the CuPc growth condition (*i.e.*, thermally evaporated in *ca.* $10^{-6}$ Torr base pressure and at low evaporation rate, *ca.* 0.3 Å/s, with the substrate kept at room temperature) agrees with an α-CuPc film. The CuPc ultrathin-film polycrystalline arrangement is demonstrated by T. Li *et al.,* using grazing incidence X-ray diffraction.[33] Figure 1a (right-hand panel) illustrates the polycrystalline configuration expected for CuPc nanometer-thick films. Notice that, at the vicinity of the Au surface, a significant portion of the CuPc crystalline domains is found at *ca.* 25-30º to the substrate,[47] whereas a small amount of molecules is arranged almost perpendicularly (*ca.* 65º).[8] As the CuPc films get thicker, the most-likely perpendicular arrangement prevails (*ca.* 65º, Figure 1a).[8]

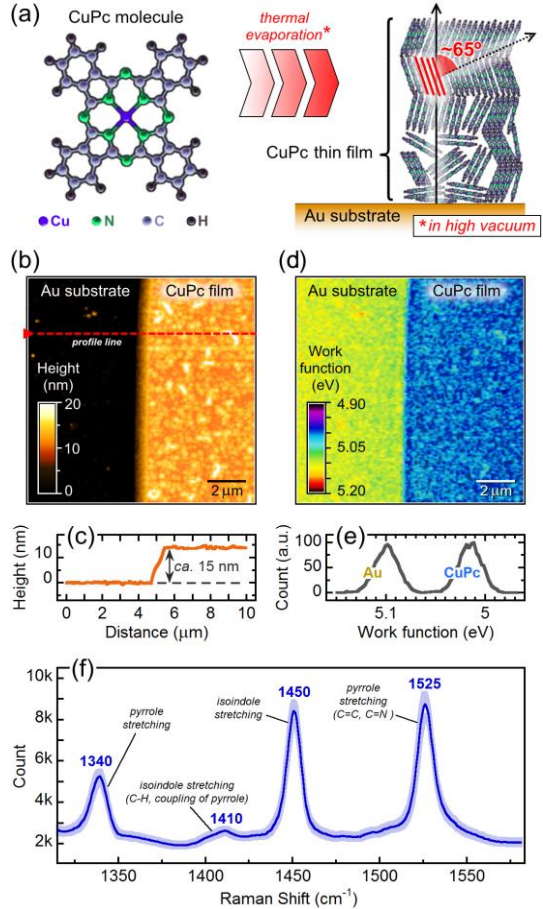

**Figure 1: Structure, morphology, and topography of CuPc nanometer-thick films. (a)** Schematic molecular packing structure of CuPc films thermally-evaporated on Au substrates. **(b)** AFM image of the CuPc film grown on the Au substrate. **(c)** CuPc film height profile. **(d)** KPFM image of an as-grown CuPc film on the Au substrate. **(e)** Work function distributions for the Au substrate and the CuPc film on it. **(f)** Raman signatures of the CuPc film ($t$ = 15 nm) on the Au substrate.

An AFM image of a *ca.* 15 nm thick CuPc film grown on the Au substrate is exhibited in Figure 1b. The size distribution of CuPc domains is comparable with the expected in the literature,[17] namely, a *ca.* 3 nm root mean square (RMS) roughness and a maximum variation of ± 5 nm over the CuPc film surface (considering a 40 μm$^2$ surface area). For the bottom electrode, we found a *ca.* 1 nm RMS roughness and a ± 2 nm maximum height variation. The corresponding height profile (dashed line in Figure 1b) is shown in Figure 1c, indicating the adequate CuPc film thickness of $t$ = (15 ± 5) nm. We also call attention that the amplitude of the half-height variation found over the CuPc-film and height variation at the bottom-electrode surface (5 and 4 nm, respectively), must be subtracted both from the 15 nm thickness in order to accurately estimate the *nCap* effective thickness, $t^{eff} \approx 6$ nm. Figure 1d exhibits the corresponding KPFM image for the CuPc deposited on the Au substrate. The work function distributions for both materials are plotted in Figure 1e. We found a *ca.* 90 meV energy difference between Au and CuPc work functions. Our measured CuPc work function is ~5 eV, which is compatible with the value obtained for metal-phthalocyanine thin films.[48] This agreement suggests that the Au/CuPc heterostructure should exhibit Ohmic charge carrier injection since holes play a major role in charge transport,[3,38,49] and no Schottky barrier for holes is expected.

Figure 1f displays the Raman spectrum acquired from the CuPc film ($t$ = 15 nm) grown on the Au substrate. The corresponding raw data are provided as *Supporting Information* (Figure S1). The literature reports that CuPc growth on crystalline substrates may occur in two film polymorphs, namely, α and β.[45] In both forms, the CuPc molecules are arranged in a slipped-stack pattern, whereas a primary difference is the angle of inclination of the molecular plane with respect to the substrate. In Figure 1f, the band assigned at 1340 cm$^{-1}$ corresponds to the pyrrole $C_\alpha - C_\beta$ stretching vibration,[50,51] whereas the band at 1525 cm$^{-1}$ corresponds to $C_\beta - C_\beta$ stretching of those pyrrole groups – the last one is a totally symmetric vibration.[50,51] The weak band at 1410 cm$^{-1}$ and the intense band at 1450 cm$^{-1}$ correspond to the isoindole stretching. Both are typical for CuPc molecular solids.[50,51] The pyrrole and isoindole bands of the CuPc thin films can change at high temperatures and electrical stimulus since they represent the vibrational motion of charge-delocalized regions of the CuPc molecule.[52,53]

The monolithically integrated device platform comprising the Au/CuPc/Au *nCap* is shown in the false-color SEM image (top-view) of Figure 2a. The device has three electrical terminals: A and A' are short-circuited by the tubular top electrode, whereas B connects to the bottom electrode. Notice that the rNM-based device platform is a three-dimensional (3D) architecture. Accordingly, a Cartesian coordinate system is displayed in each panel of Figure 2 to provide a clear visualization of the 3D device components and their position and role within the as-fabricated heterostructures. Figure 2b exhibits a lateral view of the vertical device, focusing on the active region (*i.e.,* the Au/CuPc/Au *nCap* indicated by the blue arrow). The top mechanical contact established by the rNM is expected to depend on the uniformity of the active film deposited on the bottom electrode. Consequently, knowing the active-film topography is crucial to guarantee the reproducibility of the electrical measurements. The topography of the CuPc film deposited on the Au bottom electrode, obtained by AFM, is shown in the top panel of Figure 2c. The height profile along the dotted line is plotted in the bottom panel of Figure 2c, which also evidences the arrangement of the device structure (*viz.* SiO$_2$ substrate, SiO$_2$ mesa, Au bottom electrode, and CuPc thin-film). Finally, a sketch for the cross-sectional view (*yz*- direction) of the as-fabricated *nCap* is illustrated in Figure 2d. Notice that using the $SiO_2$ mesa structure to host the bottom electrode results in an active region far from the substrate, preventing parasitic effects during further electrical evaluation. The device terminals A (common to A') and B are also illustrated in Figure 2d, electrically connected to the device top and bottom electrodes, respectively.

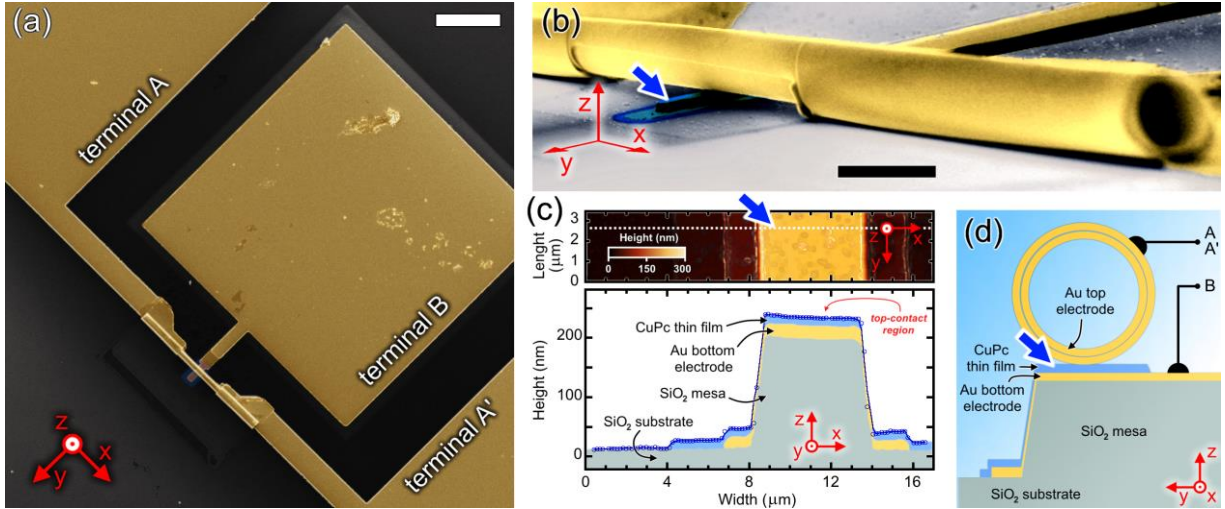

**Figure 2: Device structure and topography characterization. (a)** SEM image (top view) of the *nCap* based on rNM (scale bar corresponds to 100 μm). **(b)** SEM image (lateral view) of the *nCap* (scale bar corresponds to 10 μm). The blue arrow indicates the CuPc film active region. **(c)** AFM image exhibiting the topography of the CuPc film deposited on the bottom electrode. The height profile along the dotted line is exhibited in the bottom, highlighting $t = (15 \pm 5)$ nm. **(d)** Sketch for a cross-sectional view of the Au/CuPc/Au heterojunction with the electric terminals A, A', and B.

The electrical measurements were performed by applying a voltage *V* on the top electrode (terminal A), whereas the bottom electrode (terminal B) was grounded. Figure 3a exhibits the *I-V-T* characteristics of the CuPc films at 12 < T < 300 K. A maximum *V* of 1.5 V was applied to facilitate reproducible electrical response of the heterostructures. The detection limit of our experimental setup (10 fA) is indicated in Figure 3a. As expected for the symmetric Au electrical contacts, a negligible rectification ratio is observed within the investigated temperatures (Figure 3a). For instance, at *T* = 300 K, a rectification ratio of *ca.* 1.8 is found, whereas, at *T* = 12 K, this corresponds to *ca.* 3.7, both calculated at |*V*| = 1.5 V. Such values of rectification agree with our previous results on charge transport across thin and ultrathin CuPc molecular ensembles.[17] We attribute such a small rectification to the slight differences between interfaces since the top contact is established in an aqueous solution after the CuPc film deposition.[3] As the variation of rectification ratio does not imply any diode-like behavior, we proceed evaluating the devices for *V* > 0 without losing generality.

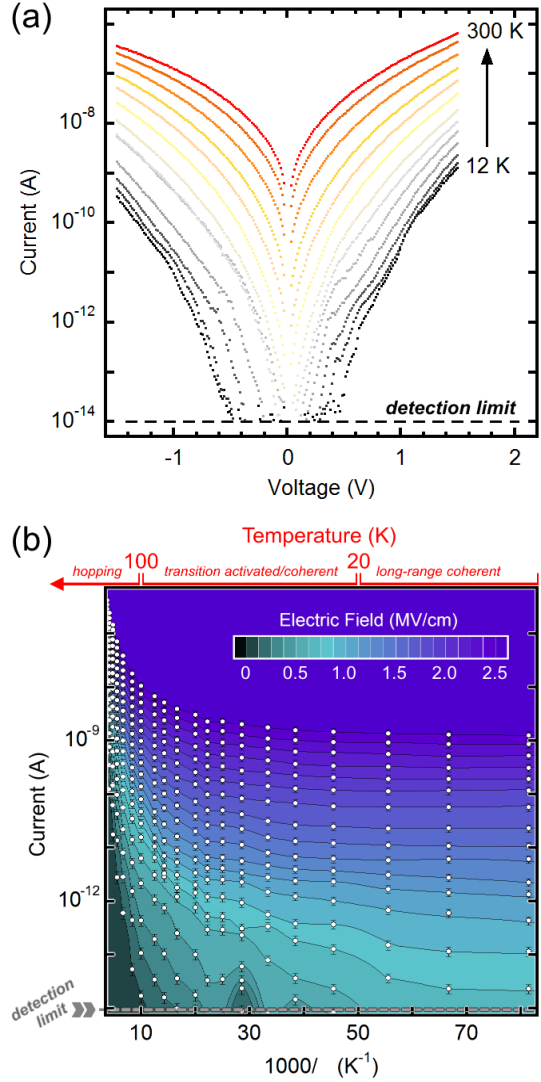

**Figure 3: *I-V-T* characteristics of the CuPc film. (a)** *I-V* traces as a function of *T* (from 12 to 300 K). **(b)** *I-V-T* data plotted in the Arrhenius form as a function of the electric field ($E = V/t^{eff}$, $t^{eff} \approx 6$ nm). In panel **(b)**, the lines are eye-guides for the average behavior obtained from 6 *nCaps*, whereas the bars are for the respective statistical deviation. The circles in (b) correspond to the data transposed from (a), for V > zero.

Figure 3b exhibits the Arrhenius plot showing the $I$-$T^{-1}$ responses measured as a function of the electric field (*E*), as shown by the black-to-blue color scale. The top horizontal axis exhibits *T* and indicates the ranges for the different charge transport mechanisms. In the range 300 > T > 100 K, we observe a strongly thermally-activated current response within a narrow region of the Arrhenius plot (at the left side of Figure 3b). Below *T* = 100 K the charge transport is found to be weakly thermally-activated. Finally, for T < 20 K the current becomes T-independent. In our previous work, we have attributed the activationless features to the occurrence of long-range coherent sequential tunneling.[17] However, it is worth mentioning that the transition from such an activationless mechanism to the T-dependent charge transport remains an open question. Figure 3b indicates that such a transition is monotonous for $100 > T > 20$ K.

We proceed to discuss the charge transport by focusing on the high-temperature regime, where $T$-dependent electron transfer kinetics is described by Marcus theory,[24,54] according to which the electron transfer rate $k_{ET}$ between two neighboring molecules is:

$$k_{ET} = A_{(T)} \exp[-\Delta G^{\ddagger} / k_B T], \qquad (Eq.\ 1)$$

where $k_B$ is the Boltzmann constant, $A_{(T)}$ is a temperature-dependent prefactor and $\Delta G^{\ddagger}$ is the free energy barrier, given by:

$$\Delta G^{\ddagger} = (\lambda + \Delta G^0)^2 / 4\lambda, \qquad (Eq.\ 2)$$

where $\lambda$ is the reorganization energy, taking into account both nuclear ("inner") and environment ("outer") relaxations, and $\Delta G^0$ is the free energy difference between the final and initial states in the electron transfer process.[24] In Eq. 1, the prefactor $A(T)$ is given by:

$$A = [2\pi H_{DA}^2 / \hbar] / [4\pi \lambda k_B T]^{1/2}, \qquad (Eq.\ 3)$$

where $H_{DA}$ is the electronic matrix element between donor and acceptor states. Considering the electron transfer between two identical molecules under an external bias, $\Delta G^0$ must take into account the potential drop between the molecules:[55]

$$\Delta G^0 = -e\, a\, E, \qquad (Eq.\ 4)$$

where $a$ is a typical intermolecular distance. Therefore, the activation energy for electron transfer becomes:

$$\Delta G^{\ddagger} = (\lambda - e\, a\, E)^2 / 4\lambda \qquad (Eq.\ 5)$$

Let us consider two identical and neighboring molecules, labeled D (donor) and A (acceptor). For zero bias, the electron transfer rates D → A and A → D will be identical, resulting in zero net current. On the opposite limit, for very large electric fields, the transfer rate D → A will be exponentially larger than A → D, so transfer rate given by Eq. 1 will be proportional to current $I$. Therefore, we can use our transport data to obtain the reorganization energy, a truly microscopic parameter. Considering that the prefactor $A_{(T)}$ (Eq. 3) is proportional to $1/T^{1/2}$, to evaluate $\lambda$ we plot in Figure 4a, for high-temperature regime (T > 200 K), the quantity $IT^{1/2}$ as a function of $1/T$ in an Arrhenius-like semi-log graph, for several values of $E$.

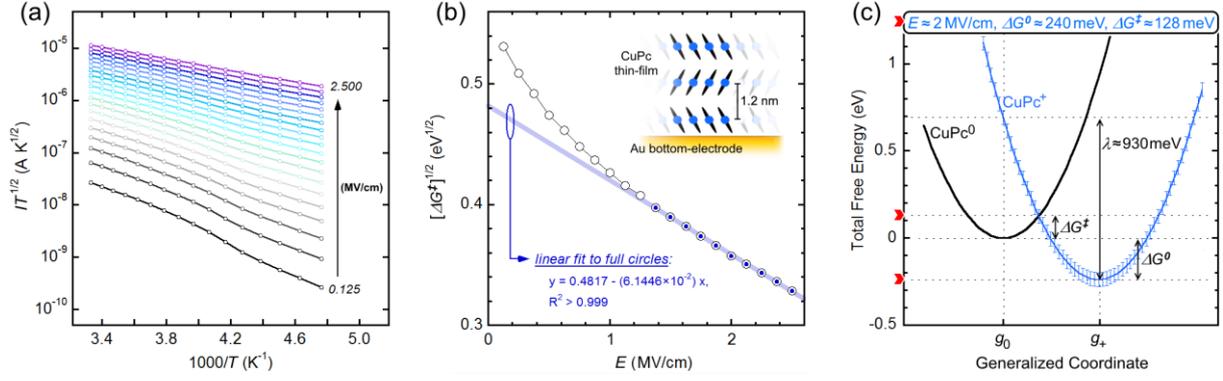

**Figure 4. Reorganization energy calculation from the experimental data. (a)** ($IT^{1/2}$) characteristics as a function of the inverse temperature for various electric fields at $200 < T < 300$ K. The circles are experimental points, whereas the lines are eye-guides for the applied electric field ($0.125 < E < 2.5$ MV/cm). **(b)** ($\Delta G^{\ddagger})^{1/2}$ *vs.* $E$ at $200 < T < 300$ K. A linear dependence on $E$ is found for $E > 1.2$ MV/cm (full circles), as demonstrated by the linear fit (blue straight line). The CuPc film structure is illustrated as inset, whereas the calculated average hopping distance is shown as *ca.* 1.2 nm. **(c)** Schematic of the total free energy curves for CuPc$^0$ and CuPc$^+$ calculated at $E = 2$ MV/cm (red arrows along *y*-axis). The reorganization energy, $\lambda$, the activation energy for hole transfer, $\Delta G^{\ddagger}$, and the free energy difference between the final (g$_+$) and initial (g$_0$) states, $\Delta G^0$, are exhibited along the potential-energy curves.

The plot shown in Figure 4a linearizes Eq. 1 and Eq. 3, allowing one to verify the charge transfer exponential-dependence on the inverse temperature, for a range of applied *E*. As a matter of fact, notice that log[ $IT^{1/2}$ ] is found to be more linearly dependent on $T^{-1}$ for larger *E*, as expected. The exponential fits to Figure 4a data returned determination coefficients ($R^2$) of *ca.* 0.993 and *ca.* 0.999, respectively for 0.125 and 2.5 MV/cm. The exponential fit parameters and the corresponding $R^2$ are provided in Table S1, whereas the adjusted curves are shown in Figure S2a, both in the *Supporting Information*. The exponential fits to Figure 4a data confirm the predictions of Marcus theory, regarding the current thermally-activated behavior and the *E*-dependence found in $\Delta G^{\ddagger}$. The overall beahavior of $\Delta G^{\ddagger}$ as a function of *E* is depicted in Figure S2b. We find that $\Delta G^{\ddagger}$ decreases from 282 meV to 108 meV as *E* increases from 0.125 MV/cm to 2.5 MV/cm. These are free energy values slightly higher than those reported using classical hopping theory.[17] Comparable energy values up to 120 meV were also calculated considering CuPc under *ca.* 1 MV/cm external electric fields, and using a molecular-dynamic/density-functional-theory hybrid model to obtain the charge carrier mobility.[55]

From $\Delta G^{\ddagger}$ obtained from the fits of Figure 4a and using Eq. 5 we can calculate the microscopic parameters $\lambda$ and *a* from the $\Delta G^{\ddagger}$ dependence on *E*. Figure 4b shows the ($\Delta G^{\ddagger})^{1/2}$ *versus E* plot, where the open circles are the experimental data and the straight line is a linear fit, which adjusts well the data for $E > 1.2$ MV/cm (smaller full circles). From the slope and intercept we find $\lambda = (930 \pm 40)$ meV and $a = (1.2 \pm 0.2)$ nm. A schematic plot of the total free energy curves for CuPc$^0$ and CuPc$^+$ states involved in the hole transfer reaction is shown in Figure 4c. The total free-energy calculation is performed for $E = 2$ MV/cm (red arrows along *y*-axis), in agreement to our proposed methodology to investigate the charge transfer rate. The obtained parameters ($\lambda$, $\Delta G^{\ddagger}$, and $\Delta G^0$) are exhibited along the potential-energy curves for the neutral and cationic molecular geometries ($g_+$ and $g_0$, respectively).

The meaured reorganization energy, $\lambda = (930 \pm 40)$ meV, is comparable with the value recently reported by S. Fatayer et al.,[32] who found $(800 \pm 200)$ meV from single-charge sensible AFM investigations in single naphthalocyanine (NPc) molecules on insulating substrates. We attribute to the presence of more phenyl groups in NPc the possible difference found between CuPc's 930 meV to NPc's 800 meV.[32] In regard to $a \approx 1.2$ nm, we call attention that such a value is very consistent with the predicted separation between the CuPc molecules in a polycrystalline arrangement (slipped-stack, $\alpha$-phase),[45,47] as illustrated in Figure 4b(inset). This result strongly suggests that the charge transport is taking place *via* a single hopping-event per molecule – *i.e.*, in a totally intermolecular fashion. This also implies the occurrence of *ca.* 5 charge-transfer events to complete the overall hopping transport across $t^{eff} \approx 6$ nm. Such findings are also in agreement with other CuPc single-molecule investigations performed by probe-tip measurements,[56,57] and also corroborates the expected hole density in the CuPc HOMO, *ca.* $10^{21}$ cm$^{-3}$.[58]

We proceed to further discuss $\lambda$, such a fundamental microscopic parameter from Marcus theory and provided by our experimental data. For this purpose, we performed *ab initio* calculations (details are provided in the Methods section) considering the charge (hole) transfer reaction between two identical CuPc molecules, as illustrated in the chemical reaction of Figure 5a. From this perspective $\lambda$ is defined as the energy cost to rearrange the CuPc molecule and its surrounding dielectric environment upon a charge transfer, and it can be calculated as proposed by Vaissier *et al*:[25]

$$\lambda = [G_0(g_+, \varepsilon_{opt}) + G_+(g_0, \varepsilon_{opt})] - [G_+(g_+, \varepsilon_{stc}) + G_0(g_0, \varepsilon_{stc})], \quad \text{(Eq. 6)}$$

where $G_0$ and $G_+$ are the energies of CuPc neutral and cationic molecules, respectively, calculated at either neutral ($g_0$) or cationic ($g_+$) geometries. Eq. 6 can be seen as the energy difference between a transition state (where charge was transferred but neither geometries or dielectric medium have yet relaxed) and the ground state (where both molecules and the medium are relaxed in their respective charge state). The CuPc static dielectric constant $\varepsilon_{stc}$ can be used to describe the environment in the initial equilibrium state, while the CuPc high-frequency optical dielectric funtion $\varepsilon_{opt}$, to describe the dielectric environment at the non-equilibrium state. Therefore, we varied independently $\varepsilon_{stc}$ and $\varepsilon_{opt}$ in a wide range to calculate $\lambda$. This analysis was performed without and with the application of an electric field of 1.0 MV/cm and the results are shown as a color map in Figure 5b,c.

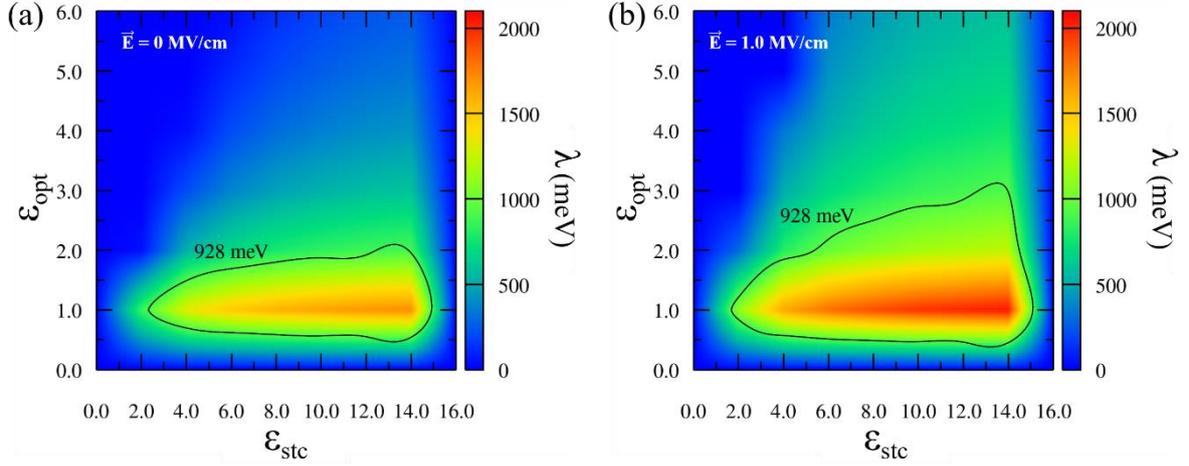

**Figure 5. CuPc reorganization energy from *ab initio* calculations. (a)** Sketch for the CuPc intermolecular hole transfer reaction. Reorganization energy (color scale) calculated for the CuPc molecule as a function of its dielectric constants ($\varepsilon_{stc}$, $\varepsilon_{opt}$), at **(b)** $E$ = zero, and **(c)** $E$ = 1 MV/cm. The calculations are for $E$ applied parallel to the *nCap*'s bottom electrode. The contour line (in black) shows the experimental value for reorganization energy between CuPc molecules, $\lambda = (930 \pm 40)$ meV, obtained in this work.

In Figure 5b,c, for both cases ($E$ = zero and $E$ = 1 MV/cm) the experimental value $\lambda = (930 \pm 40)$ meV is denoted by a black contour line. In addition, the horizontal and vertical strips indicate the experimental values for optical ($\varepsilon_{opt} = 3.50 \pm 0.25$)[59] and static ($\varepsilon_{stc} = 7 \pm 1$)[60] dielectric constants, respectively, delimited by dashed lines. The two strips intercept in a rectangular region indicating the possible range of experimental values of the two dielctric constants. Interestingly, this rectangular region is also crossed by the experimental contour line for $\lambda$, indicating a consistent agreement between theory and experiment.

## Conclusion

The demonstration of a monolithically-integrated nanodevice platform able to assess molecular reorganization energy is successfully completed. This is achieved thanks to the rNM technology, which allowed us to measure the CuPc charge-transport dependence on both temperature and electric field at the molecular level. Thereby, we calculated the CuPc reorganization energy, $\lambda = (930 \pm 40)$ meV, from studying the hole transfer reactions measured in a *ca.* 6 nm thick Au/CuPc/Au nanocapacitor. AFM, KPFM, Raman spectroscopy, and SEM attested the materials' and device's reproducibility, whereas DFT calculations provided the atomistic picture to support our experimental findings. Our approach delivers a consistent route towards electron transfer reaction characterization from charge transport measurements using monolithically-integrated electrochemical nanodevices.

## Acknowledgments

We acknowledge FAPESP (18/18136-0, 17/21117-4, 19/10188-3, 14/25979-2), CNPq (408770/2018-0), and CAPES for financial support. C.C.B.B. is a productivity research fellow from CNPq (305305/2016-6). C.C.B.B also acknowledges the support of CNPq and FAPESP (Brazil) through Inomat, INCT (CNPq 465452/2014-0 and FAPESP 14/50906-9). G.C. gratefully acknowledges the computational support of Núcleo Avançado de Computação de Alto Desempenho (NACAD/COPPE/UFRJ) and Sistema Nacional de Processamento de Alto Desempenho (SINAPAD), and FAPERJ (Processo E-26/200.008/2020) for financial support. We also thank LNNano/CNPEM: Electron Microscopy and Scanning Probe Force Microscopy laboratories. We are grateful to Davi H. S. de Camargo and Leirson D. Palermo for the technical support with clean-room processes and electrical measurement facilities, respectively. We are grateful to Carlos A. R. Costa and Otavio Berenguel for the technical support with the KPFM and Raman-spectroscopy data acquisition, respectively.

## Author information

**Corresponding Author**

*e-mail: carlos.bufon@Lnnano.cnpem.br

**Author Contributions**

L.M. fabricated the molecular transport-junction devices, carried out electrical measurements, AFM and Raman experiments, as well as performed data evaluation, wrote and revised the manuscript. G.C. performed the *ab initio* calculations and discussed the corresponding results. L.M.M.F. analyzed Raman spectra and revised the manuscript. A.B. analyzed and discussed Raman spectra. C.V.S.B. analyzed KPFM data. A.N. discussed the electrical measurements and revised the manuscript. A.R.J. discussed results and revised the manuscript. R.B.C. supervised the a*b initio* calculations, discussed the results and revised the manuscript. C.C.B.B. discussed results, revised the manuscript, and led the work.

[FIGURE]

**Table-of-content graph:**

Reorganization energy from charge transport measurements in a monolithically-integrated molecular device